# PROFILING AND VARIATION OF LASER PULSE PARAMETERS AS A WAY TO PRESERVE THE STABILITY OF SELF-INJECTED BUNCHES DURING EXCITATION OF A WAKEFIELD IN PLASMA

D. S. Bondar[1,2], V. I. Maslov[1], and I. N. Onishchenko[1]

[1]National Science Center "Kharkiv Institute of Physics and Technology",
[2]Karazin Kharkiv National University, Kharkiv, Ukraine
E-mail: bondar.ds@yahoo.com

The paper considers the excitation of a wakefield in a metal-density plasma using a chain of x-ray laser pulses. The profiling parameters and the necessary parameters of laser pulses for obtaining stable high-quality bunches are found. An essential problem is the destruction of self-injected bunches in the course of their motion. The results of the study are one of the ways to solve the problem of transverse betatron oscillations, which lead to the destruction of bunches.

## INTRODUCTION

The field of laser-plasma acceleration has seen significant advancements over the past few decades. The concept of using ultra-high intensity laser pulses to achieve extreme material states in the laboratory has become almost routine with the development of petawatt and even zettawatt-exawatt class lasers [1-3]. These lasers have been constructed for specific research activities, including particle acceleration, inertial confinement fusion, and radiation therapy, and have been used to generate secondary sources such as x-rays, electrons, protons, neutrons, and ions [4-6].

One of the key areas of research in this field is the excitation of a wakefield in a plasma. The use of high-density plasma in the excitation of a wakefield has been shown to produce high-quality electron beams [7-9]. This is particularly relevant in the context of laser-plasma accelerators, where the production of monoenergetic electron beams is a key goal [10-12].

The use of laser pulse chains in this context has also been explored. The ability to produce high intensities with these lasers opens up new possibilities for the study of relativistic and even nuclear physics [13-15]. Furthermore, these lasers can be synchronized with other large-scale facilities, such as megajoule lasers, ion and electron accelerators, and x-ray sources, further expanding their potential applications [16].

One of the key challenges in this field is the preservation of the stability of self-injected bunches during their motion. This is where the concept of laser profiling comes into play. Laser profiling allows for the control of the injection and acceleration of electrons in plasma wakefield, potentially solving the problem of transverse betatron oscillations that can lead to the destruction of bunches.

The potential of laser-plasma accelerators in reproducing space radiation in the laboratory has also been demonstrated, further expanding the applications of this technology. Comprehensive studies on the structure and dynamics of laser wakefield acceleration, including the measurement of the electron beam, have provided crucial insights for understanding and improving the stability of self-injected bunches. The exploration of the highly nonlinear broken-wave regime of laser wakefield acceleration has provided valuable insights into the complex dynamics involved in this process [17-21].

The use of laser-plasma acceleration, high-density plasma in the excitation of a wakefield, laser pulse chains, and laser profiling present a promising approach to preserving the stability of self-injected bunches during excitation of a wakefield in plasma.

The paper investigates the issue of profiling laser pulses and the effect of this process on self-injected bunches. The authors have improved the quality of self-injected bunches by profiling the laser sequence. In addition, suppression of the transverse instability (expansion) of bunches is considered by the method of parameter variation using numerical simulation.

## 1. STATEMENT OF THE PROBLEM

The process of using a laser pulse to excite a wakefield in a plasma is investigated by numerical simulation. The density of this plasma is approximately equal to the density of free electrons in metals, with a focus on profiled bunches. A profiled sequence of laser pulses located at a distance of one plasma wavelength is considered. In addition, suppression of the transverse expansion (instability) of bunches by varying the parameters is demonstrated.

The main tasks were:
1. Show the advantages of a profiled sequence compared to a non-profiled one.
2. Consider a way to prevent bunch decay by varying parameters.

Main system parameters include: an unperturbed plasma electron density, normalized on the graph to $n_{0e}=10^{23}$ cm$^{-3}$, a ratio between the plasma frequency and laser frequency, $\omega_{pe}/\omega_0$, equating to 0.1008 (where $\omega_0$ and $\omega_{pe}$ are the laser and plasma frequencies respectively). The laser wavelength was $\lambda_l=10.6$ nm, and all lengths, distances and coordinates were normalized to this laser wavelength, $\lambda_l$. The laser pulse was directed along the system's axis. The plasma wavelength is $\lambda_{pe}=106$ nm.

The dimensions of the simulation window were a length of 800 and width of 50. The laser amplitude, a, was normalized to the overturning field proportional, $E_0$

($a=EE_0^{-1}$), where $E_0=m_ec\omega_0/e$. The ratio of ion to electron mass was determined to be 1836. Time normalized to the electromagnetic wave period $T_0=0.352$ fs. At consideration of the influence of the profiling effect on self-injected bunches the sequences of laser pulses were used.

First **profiled sequence S-P**: 1st pulse (half-length 2, half-width 1.5, amplitude 2); 2nd pulse (half-length 1, half-width 3, amplitude 4).

In addition, in order to carry out a comparative analysis, an **unprofiled sequence S-UnP** with two identical laser pulses was considered: half-length 2, half-width 1.5, amplitude 2.

In total, **a comparison was studied**, which gives a comprehensive picture of the impact of profiling effects.

## 2. RESULTS OF SIMULATION

### 2.1. INFLUENCE OF THE EFFECT OF LASER PULSE SEQUENCE PROFILING ON SELF-INJECTED BUNCHES

Figures 1-4 show the plasma density distribution as a function of the transverse and longitudinal coordinates for a profiled sequence of bunches, longitudinal momentum distribution and its distribution function. Time points 175 and 200 are shown in normalized units. At these moments, the self-injected bunch moves through wake bubble. The figures show the optimal moment of the wake process, when it is supposed to use a self-injected bunch in practice.

Figures 5-7 illustrate the distributions in the case of a unprofiled sequence. Comparison of Figures 3 and 4 allows us to show the dynamics at short times for the case of a profiled sequence, which, according to research results, is optimal.

From Fig. 1, 2 one can see that the scatter of the longitudinal momentum values is insignificant during the motion of the self-injected bunch. This is the merit of profiling. The bunch retains its structure, the head and the central part of the bunch do not disintegrate. Also, no significant transverse instability of bunches is observed.

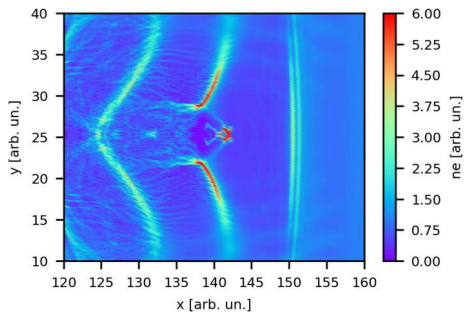

*Fig. 1. Plasma electron density distribution $n_e(x, y)$, $t=175T_0$. Profiled sequence S-P.*

A comparison of Figures 3 and 4 shows that even in a short time (25 laser periods) in the case of a profiled sequence, the increase in the maximum peak of the longitudinal pulse values is 12.5%.

In this case, an increase in the number of particles with this energy is observed by approximately 43% in the region of the main peak. This clearly indicates the advantage of profiling in the form of increasing the monoenergeticity of self-injected bunches.

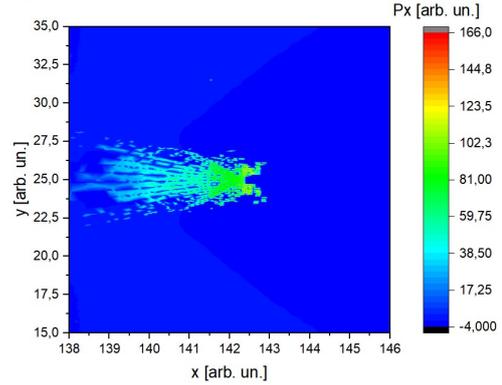

*Fig. 2. Longitudinal momentum distribution $P_x(x, y)$, $t=175T_0$. Profiled sequence S-P.*

Let us perform a direct comparison of bunches in the cases of profiled and non-profiled sequences. Comparison of data in Fig. 2 and fig. 5 allows to conclude that in the case of a profiled sequence, more stable bunches are observed.

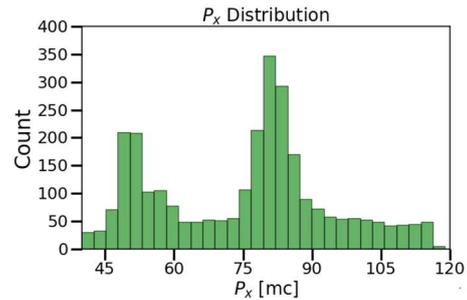

*Fig. 3. $P_x$ distribution function, $t=175T_0$. Profiled sequence S-P.*

In the absence of profiling (Fig. 5), the structure of the bunch is preserved, but a large number (up to 30% of the volume) of electrons is lost in the plasma. This is a significant disadvantage. Therefore, the case of profiling has an advantage in the context of maintaining bunch stability.

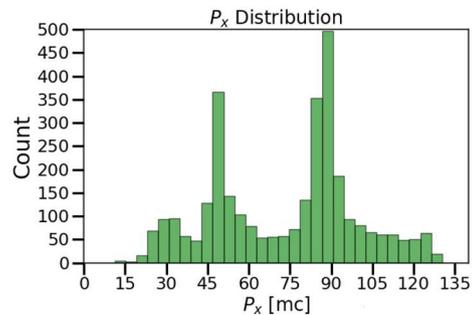

*Fig. 4. $P_x$ distribution function, $t=200T_0$. Profiled sequence S-P.*

In a number of cases, self-injection is not observed at all. This circumstance, taking into account the characteristics of these laser sequences, explains the advantage of laser pulse profiling in cases where self-injection as such is absent without profiling. Figure 6 indicates that the non-profiled sequence has a wide energy spread relative to the profiled case and a small number of high-energy electrons.

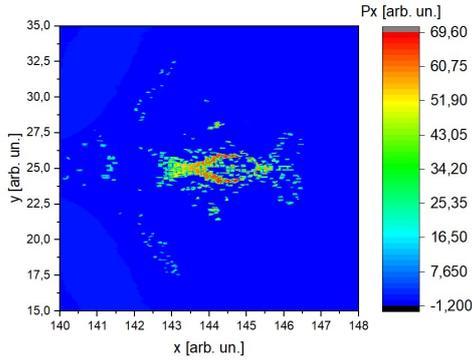

*Fig. 5. Longitudinal momentum distribution $P_x(x, y)$, $t=175T_0$. Unprofiled sequence S-UnP.*

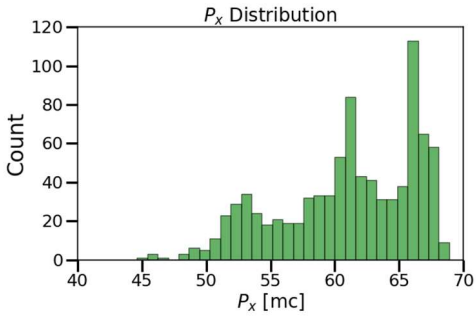

*Fig. 6. $P_x$ distribution function, $t=175T_0$. Unprofiled sequence S-UnP.*

In addition to everything said earlier, the effect of saving a structure of self-injected bunch during its movement in plasma was observed for a longer time than in the non-profiled case. Namely, it was observed that at time $t=375T_0$ in the profiled case the self-injected bunch structure was saved. At the same time, in a non-profiled case, at this moment the bunch is completely destroyed.

Based on the results of the study, the advantages of the profiled sequence are obvious.

## 2.2. SUPPRESSION OF THE TRANSVERSE SPATIAL EXPANSION OF SELF-INJECTED BUNCHES (INSTABILITY) BY VARYING THE PARAMETERS OF LASER PULSES

Laser pulse profiling is intended for fine tuning of self-injected bunches, increasing their momentum, monoenergeticity. In this case, each of the pulses has its own parameters. And the calculation of profiling is often far from simple and obvious.

At the same time, in a large number of cases, it is required to keep the parameters of the bunches of the sequence the same. At the same time, it is required to suppress the transverse instability of electron bunches, in fact, the separation of the bunch into two parts, which scatter along the radius from the axis of the system.

The most difficult case is when, in the case of a chain of two laser pulses, it is required to suppress the transverse expansion (instability) of only one bunch.

In this section, it is supposed to investigate the suppression of this instability using numerical simulation by the method of parameter variation. The reason for this is that a simultaneous change in the parameters of laser pulses (even a slight one) affects several self-injected bunches simultaneously.

The basic sequence of laser pulses that was considered consisted of two identical laser pulses, which are located at a distance of one plasma wavelength. Let's call this sequence **R-2** (half-length 1, half-width 3, amplitude 5).

Fig. 7, 8 show the simulation results for the R-2 sequence. At first glance, the problem becomes obvious, the decay of the first self-injected bunch into three parts. Two parts of the bunch scatter along the axis and the bunch ceases to exist as a whole.

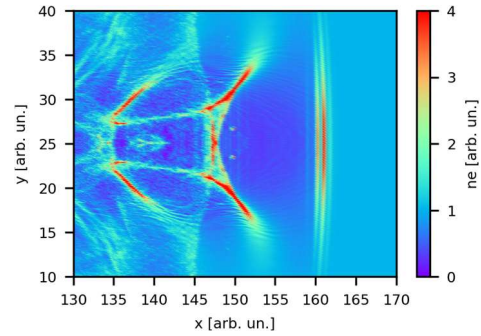

*Fig. 7. Plasma electron density distribution $n_e(x, y)$, $t=175T_0$. Sequence R-2.*

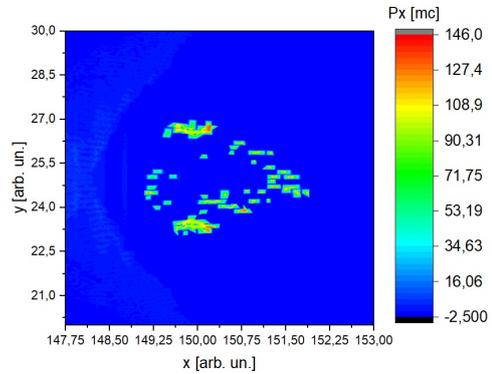

*Fig. 8. Longitudinal momentum distribution $P_x(x, y)$, $t=175T_0$. Sequence R-2.*

The sequence of two pulses, which we will call **R-2M1** (half-length 1.5, half-width 2, amplitude 5), is the result of parameter variation, thanks to which it was possible to suppress the instability of the first bunch, while preserving the second one. Fig. 9, 10 show the distribution of the first self-injected bunch for sequence R-2M1, after applying the parameter variation method. Fig. 11, 12 show the simulation result for sequence **R-2M2** (half-length 1.5, half-width 2, amplitude 4.75). The results of the study for sequence R-2M2 allow us to conclude that using the parameter variation method, even without changing the amplitude, it is possible to achieve the cessation of the transverse decay of bunches, stabilization of the bunch, and an increase in its average energy (longitudinal momentum). A slight additional amplitude variation contributes to the more monoenergetic bunch. At the same time, a significant variation in the amplitude value leads to the disappearance of the bunch.

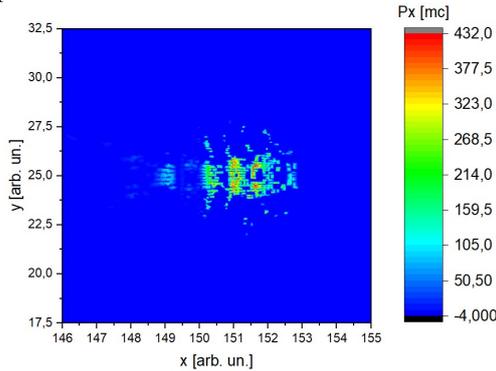

*Fig. 9. Longitudinal momentum distribution $P_x(x, y)$, $t=175T_0$. Sequence R-2M1.*

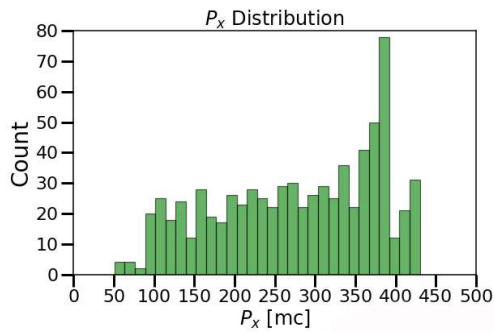

*Fig. 10. $P_x$ distribution function, $t=175T_0$. Sequence R-2M1.*

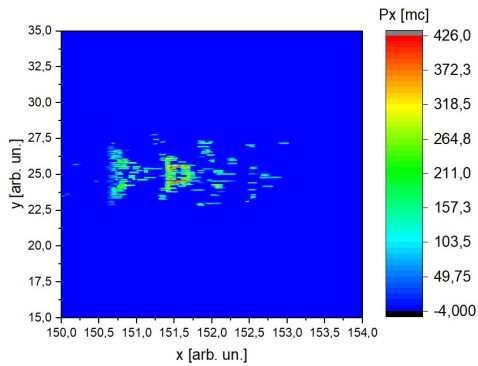

*Fig. 11. Longitudinal momentum distribution $P_x(x, y)$, $t=175T_0$. Sequence R-2M2.*

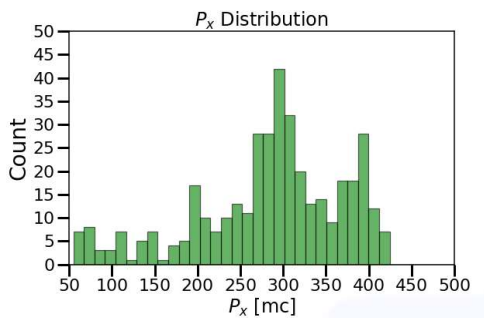

*Fig. 12. $P_x$ distribution function, $t=175T_0$. Sequence R-2M2.*

## CONCLUSIONS

Based on the results of the study, it can be concluded that profiling has clear advantages over non-profiled sequences. Profiling promotes the formation of more high-energy and monoenergetic bunches, compared to non-profiled cases. A number of comparisons were made between profiled sequences and non-profiled sequences. At the same time, the obvious advantages of profiling are emphasized. In addition, a method was considered for suppressing the transverse decay of bunches by varying the parameters. Parameters were found at which it is possible to suppress the decay of one bunch, while saving the second one (without a significant deterioration in its characteristics and with saving second bunch structure).

## ACKNOWLEDGMENTS

This work is supported by National Research Foundation of Ukraine "Leading and Young Scientists Research Support", grant agreement # 2020.02/0299.

## ПРОФІЛЮВАННЯ ТА ВАРІЮВАННЯ ПАРАМЕТРІВ ЛАЗЕРНОГО ІМПУЛЬСУ ЯК ШЛЯХ ЗБЕРЕЖЕННЯ СТАБІЛЬНОСТІ САМОІНЖЕКТУВАНИХ ЗГУСТКІВ ПІД ЧАС ЗБУДЖЕННЯ КІЛЬВАРТНОГО ПОЛЯ В ПЛАЗМІ

*Д.С. Бондар, В.І. Маслов, І.М. Оніщенко*

У статті розглядається збудження кільватерного поля в плазмі металевої густини за допомогою ланцюжка імпульсів рентгенівського лазера. Знайдено параметри профілювання та необхідні параметри лазерних імпульсів для отримання стабільних високоякісних згустків. Суттєвою проблемою є руйнування самоінжектованих згустків під час їх руху. Результати дослідження є одним із шляхів вирішення проблеми поперечних бетатронних коливань, які призводять до руйнування згустків.